\documentclass[pdflatex,sn-vancouver-num]{sn-jnl}

\usepackage{graphicx}%
\usepackage{multirow}%
\usepackage{amsmath,amssymb,amsfonts}%
\usepackage{amsthm}%
\usepackage{mathrsfs}%
\usepackage[title]{appendix}%
\usepackage{xcolor}%
\usepackage{textcomp}%
\usepackage{manyfoot}%
\usepackage{booktabs}%
\usepackage{algorithm}%
\usepackage{algorithmicx}%
\usepackage{algpseudocode}%
\usepackage{listings}%
\usepackage[normalem]{ulem}
\usepackage{comment}

\begin{document}
\title{Memory-driven topological ordering during the transition from dormant to migrating epithelia}

\author[1]{Richard Ho}
\equalcont{These authors contributed equally to this work.}

\author[2]{Anna Lång}
\equalcont{These authors contributed equally to this work.}

\author[2]{Emma Lång}
\equalcont{These authors contributed equally to this work.}

\author*[2]{Stig Ove Bøe}\email{s.o.boe@medisin.uio.no}

\author*[1]{Luiza Angheluta}\email{luiza.angheluta@fys.uio.no}

\affil[1]{Department of Physics, University of Oslo, Sem S$\textrm{\ae}$landsvei 24, 0371, Oslo, Norway}

\affil[2]{Department of Microbiology, Oslo University Hospital, Forskningsveien 1, 0373, Oslo, Norway}

\abstract{
Transitions from quiescence to collective migration in epithelia underlie wound healing and cancer invasion, yet their physical origin remains poorly understood. Here we show that quiescent epithelial monolayers store spatially contractile stresses that function as a form of mechanical memory. Upon serum-induced reactivation, these pre-stressed regions nucleate extensile asters that emit propagating polarity domain walls. Along these interfaces, topological defects are created, advected and annihilated, leading to defect coarsening with faster kinetics than by elastic interactions. An active elastic model quantitatively reproduces the observed dynamics and identifies stored stress as the origin of rapid topological reorganization. Our results establish a mechanism in which mechanical memory in quiescent epithelia triggers active stress release, driving collective migration via rapid topological ordering, distinct from conventional unjamming and flocking transitions.
}


\maketitle 

\section{Introduction}

Epithelial tissues can transition from a quiescent state, characterized by limited cell division and motility, to coherent, tissue-scale migration with well-defined polarity. This transition underlies fundamental biological processes such as morphogenesis, wound healing, and cancer invasion~\cite{cheung2025collective}. Despite extensive experimental and theoretical work, the principal mechanisms governing the transition of an initially static confluent epithelial monolayer with disorganized migration patterns to a state of collective migration remain incompletely understood.

A widely used theoretical framework for collective motion is provided by polar active matter, in particular the Toner-Tu hydrodynamic theory~\cite{toner1995long,toner1998flocks}. In this description, self-propelled units interact via local alignment and move in a momentum-dissipating medium, giving rise to a flocking transition from disordered to ordered motion. This framework assumes fluid-like rearrangements and rapid stress relaxation, such that large-scale polar order emerges through a coarsening dynamics governed by the annihilation of topological defects. The phase-ordering kinetics in such active fluids is characterized by an algebraic decay of defect density $N_d(t)\sim t^{-1}$~\cite{rana2020coarsening,chardac2021topology,mondal2025coarsening}. However, compressibility and advective nonlinearities in Toner-Tu systems can stabilize long-lived defect structures such as asters, leading to slow or arrested coarsening and glassy dynamics~\cite{mondal2025coarsening}.

An alternative perspective is provided by the jamming–unjamming paradigm, in which densely packed cells transition from a solid-like, caged state to a fluid-like, motile phase~\cite{angelini2010cell,angelini2011glass,szabo2006phase,park2015unjamming}. Although this approach captures changes in mobility and mechanical rigidity associated with epithelial-to-mesenchymal transitions, it assumes that collective motion arises from enhanced cellular rearrangements, i.e. fluidization. However, recent experiments using confluent Madin-Darby canine kidney (MDCK) cells have reported coherent flows and dynamic heterogeneity in states that remain mechanically solid-like, suggesting that fluidization alone is insufficient to explain the onset of collective migration~\cite{shen2026dynamic}. Moreover, serum-induced flocking in the immortalized human keratinocyte cell line HaCaT occurs in the absence of cell rearrangements, demonstrating that local neighbor exchanges are not required to initiate collective epithelial migration~\cite{laang2024topology, shen2025flocking}.

Recent studies have emphasized the role of topological organization in epithelial velocity fields, where integer-charged defects structure local flow in macroscopic aster- and vortex-like patterns~\cite{laang2024topology,angheluta2025topological}. During the onset of collective migration, these defects undergo annihilation events that reorganize the velocity field and contribute to the emergence of long-range polar order~\cite{laang2024topology}. Although this picture shares features with defect-mediated ordering in active polar fluids~\cite{chardac2021topology,rana2020coarsening}, the physical mechanisms that govern defect motion and the resulting coarsening kinetics in epithelial tissues remain largely unresolved. In addition, the physical mechanisms responsible for initiating defect-mediated coarsening in epithelia have not been defined.

An open question is whether collective migration emerges from the coordination of the intrinsic cell motility alone, after stimulation, or whether it is also guided by the mechanical state of the tissue prior to activation. Even in the absence of cell movement, epithelial monolayers sustain heterogeneous actomyosin-generated stress fields, suggesting that quiescent tissues may store spatially heterogeneous stress patterns~\cite{laang2022mechanical}. 

Here we demonstrate that the activation of collective migration in confluent monolayers of HaCaT keratinocytes is governed by heterogeneous contractile stresses. The stresses are formed and kept stable during serum-free conditions that favor quiescence, and rapidly reorganize into extensile asters after serum activation, which promote quiescence exit. The asters have a direct role in forming propagating polarity domain walls for which defects of opposite charges are nucleated, transported, and annihilated in pairs, giving rise to defect coarsening kinetics that is faster than expected from pairwise defect interactions alone.

To explain the experimental observations, we introduce an active elastic model that couples tissue deformation to actomyosin-generated stress. The model quantitatively reproduces the experimentally observed defect kinetics and identifies metastable actomyosin-induced asters as the source of advective defect transport and the resulting accelerated coarsening. Our results establish a mechanism for collective migration in which internally stored mechanical stresses guide the emergence of large-scale polarity, providing a route to flocking in epithelial tissues that is distinct from both classical jamming–unjamming scenarios and active polar fluid theories.

\begin{figure}[t!]
    \centering \includegraphics[width=\linewidth]{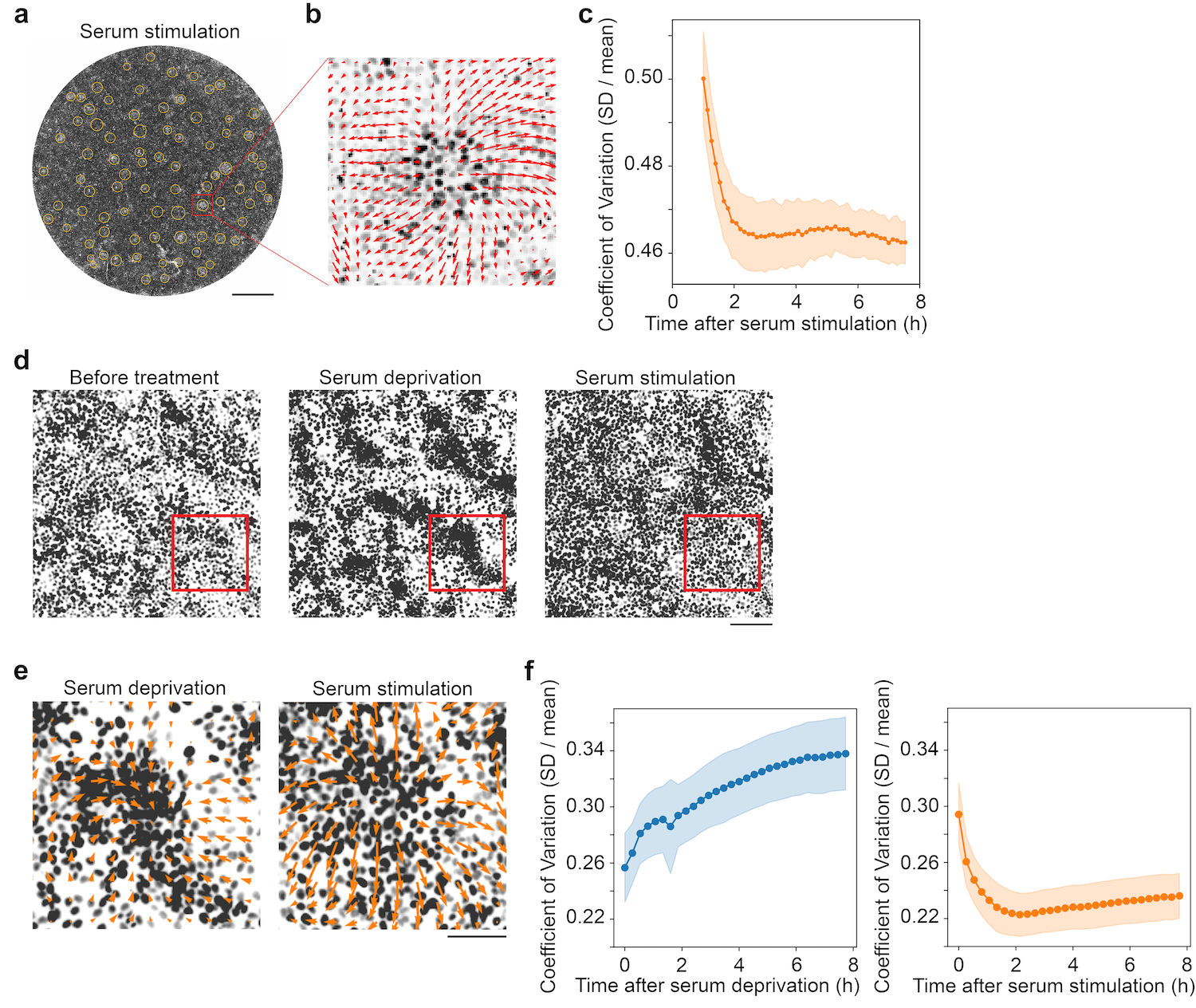}
\caption{\textbf{Heterogeneous cell densities controlled by serum-born components}. {\bf{a}}, Snapshot of HaCaT monolayer 1 h after serum stimulation. Expanding clusters of cells, formed during three days of serum deprivation, are highlighted by circles. Scale bar, $1\,\ mm$. {\bf{b}}, PIV showing velocity profile around a single expansion center (highlighted with red box in {\bf{a}}) 1.1 h after serum stimulation. {\bf{c}}, Coefficient of variation showing cell density as function of time after serum stimulation. {\bf{d}}, Snapshot showing mCherry-tagged Histone H2B before treatment (right panel), 4.0 h after serum deprivation (middle panel), and 1.3 h after serum stimulation (right panel). Scale bar, $200\,\mu m$. {\bf{e}}, PIV showing velocity profile that contract 1.3 h after serum deprivation (left panel) and expand 0.5 h after serum stimulation (right panel). Zoomed in area is highlighted with red boxes in {\bf{d}}. Scale bar, $100\,\mu m$. {\bf{f}}, Coefficient of variation showing cell density as function of time after serum deprivation (left panel) and serum stimulation (right panel).}
    \label{fig:fig1}
\end{figure}
\section{Results}

\subsection{Heterogeneous cell densities controlled by serum-born components}

We use a well-established experimental system based on the immortalized HaCaT keratinocyte cell line~\cite{laang2018coordinated,laang2022mechanical,laang2024topology,abdo2026novo,shen2025flocking}. A confluent monolayer is rendered quiescent by serum deprivation and subsequently stimulated by serum re-exposure to trigger collective migration. The duration of serum deprivation is a critical parameter, as it allows the accumulation of internally stored mechanical stresses that are rapidly amplified upon serum stimulation and contribute to the onset of flocking~\cite{laang2022mechanical}. Within $\sim 15$ h after stimulation, the tissue develops quasi–long-range polar order in the migration direction on millimeter scales.

To characterize the mechanical state underlying this transition, we quantify spatiotemporal variations in cell density using mCherry‑tagged Histone H2B, which labels cell nuclei. Figure~\ref{fig:fig1}a shows that HaCaT monolayers that have been subjected to serum deprivation contain multiple local contractile centers, visible as clusters of elevated cell density distributed throughout the monolayer. Immediately following serum stimulation these sites act as nucleation points for radial migration that organize the surrounding tissue into locally coherent expansion patterns (Supplementary video 1). This is further illustrated in Fig.~\ref{fig:fig1}b, which shows the cell velocity field surrounding a representative expansion center, where cells move radially outward in an aster-like configuration (Supplementary video 2). The emergence of these migration patterns is accompanied by a reduction in cell density fluctuations. As shown in Fig.~\ref{fig:fig1}c, the coefficient of variation of cell density decreases following serum stimulation, indicating that initially heterogeneous density fluctuations are gradually relaxed as expansion centers reorganize the tissue.

To better understand the role of serum deprivation in forming contractile sites in the monolayer, we study the cell density distribution on short timescales after serum withdrawal and serum stimulation. We observe that the cell monolayer responds immediately to serum deprivation by reorganizing into a heterogeneous cell density pattern with local sites of high cell density (Fig.~\ref{fig:fig1}d and Supplementary video 3). Between 1 and 2 h after serum depletion, these contractile sites develop inward migration patterns, as shown in the velocity field in Fig.~\ref{fig:fig1}e (left panel). Upon subsequent serum stimulation, the high-density regions, which formed during the serum-depletion period, rapidly invert their mechanical character to form outward-flowing asters, indicating a rapid change in the underlying stress state (Fig.~\ref{fig:fig1}d and e, Supplementary video 3). This inversion is accompanied by a corresponding reduction in cell density heterogeneity. Figure~\ref{fig:fig1}f shows that the coefficient of variation initially increases during serum deprivation, reflecting the formation of contractile sites (Fig. \ref{fig:fig1}f, left panel), and then decreases markedly after stimulation as these structures relax and expand (Fig. \ref{fig:fig1}f, right panel).

The rapid re-distribution of cell density in response to serum depletion and serum stimulation demonstrate that serum-born components act as a molecular switch between distinct tissue states. Serum deprivation induces the formation of localized contractile sites leading to a heterogeneous cell density, whereas serum stimulation rapidly converts these sites into extensile expansion centers that drive large-scale reorganization of the epithelial monolayer towards a more uniform cell density distribution.

\begin{figure}[t!]
    \centering 
    \includegraphics[width=0.9\linewidth]{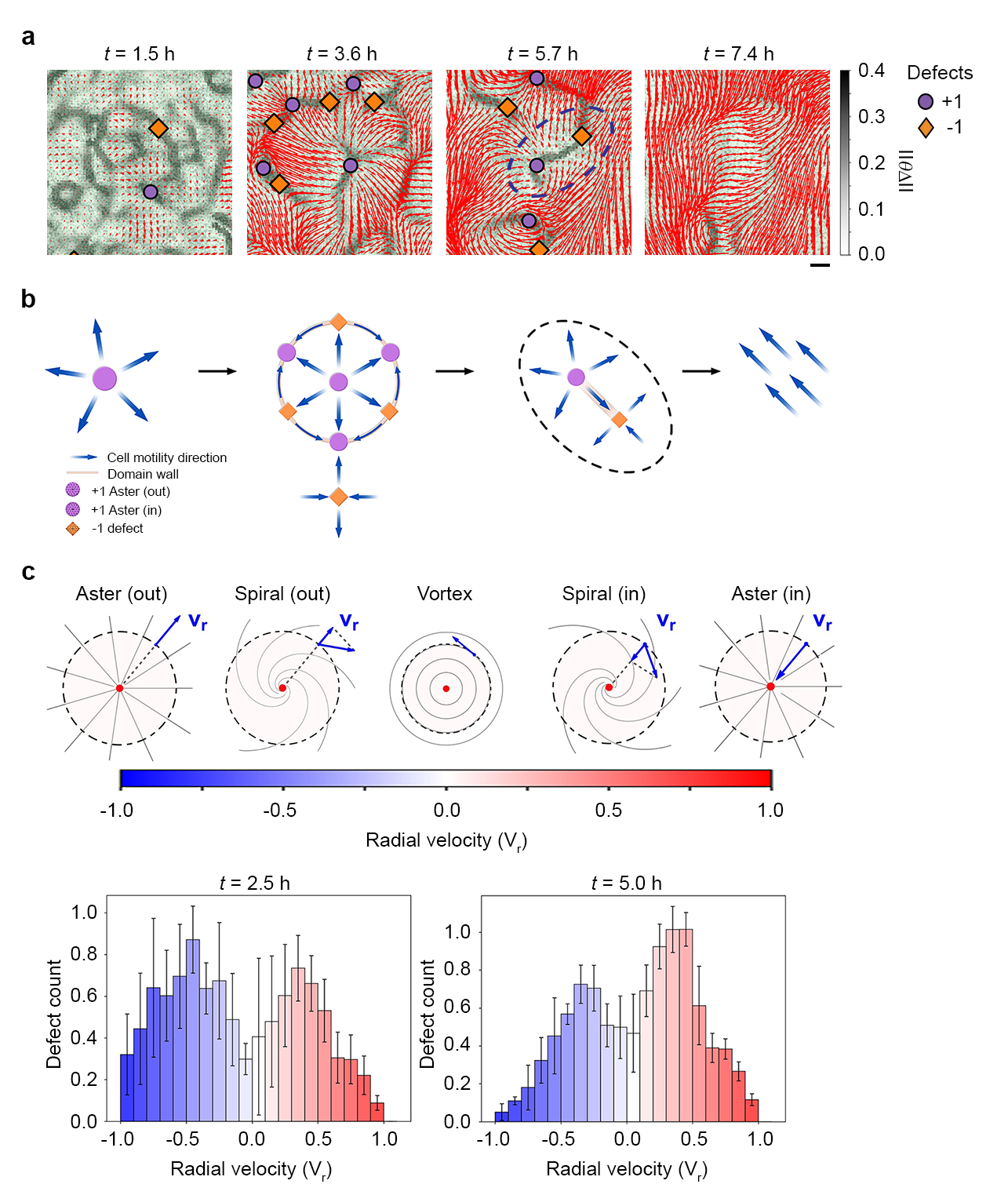}
    \caption{\textbf{Evolution of topological defects and domain walls originating from extensile cell clusters}. {\bf{a}}, Snapshots of the velocity profile for an outward aster originating from a high-density cell cluster. Topological defect positions are labeled as purple circles (+1 defect) and orange diamonds (-1 defects). Domain walls are shown in gray scale. The velocity field is represented by red arrows. Scale bar, $100\,\mu m$. {\bf{b}}, Illustration of the different stages in the lifetime of an initial outward aster formed at the onset of serum stimulation. {\bf{c}}, Probability distribution of different types of $+1$ defects at early time ($\sim 2.5$ h) versus late time ($\sim 5$ h) after serum activation. The color bar displays the radial velocity.}
    \label{fig:fig2}
\end{figure}

\subsection{Serum-induced topological defects, domain walls, and collective migration}

We next examine how the mechanically pre-patterned compressive sites induced by serum deprivation control the emergence of collective motion after serum stimulation. To this end, we track the evolution of local velocity fields and their associated topological structures. Topological defects with charge $s=\pm1$ are identified as phase singularities in the velocity orientation field, corresponding to zeros of the velocity vector. These defect cores are detected and tracked using the Halperin–Mazenko formalism based on the determinant of the velocity gradient tensor, $D = \det(\nabla \mathbf{V})$~\cite{angheluta2025topological,skogvoll2023unified}. In parallel, we identify polarity domain walls as regions of large orientation gradients, quantified by $\|\nabla \theta\|$, where $\theta = \arg(\mathbf{V})$.

Figure~\ref{fig:fig2}a shows representative experimental snapshots of the early post-stimulation dynamics, with the corresponding schematic evolution illustrated in Fig.~\ref{fig:fig2}b (Supplementary video 4). An initially localized outward aster emerges at a former compressive site and acts as a transient source of persistent radial motion. The background field in Fig.~\ref{fig:fig2}a, which represents the magnitude of the orientation gradient, reveals the spontaneous emergence of sharp interfaces separating regions of distinct migration direction. These interfaces constitute polarity domain walls that nucleate at the periphery of the expanding aster and partition the surrounding migration into domains of coherent motion. The domain walls also act as sources of defect nucleation. As the dynamics unfold, a domain wall extends from the expanding aster and connects to a nearby $-1$ defect. This connection mediates an annihilation event that removes the aster and relaxes the local migration into a coherently ordered patch. This sequence of events directly link mechanically pre-conditioned expansion centers to domain wall nucleation and defect-mediated coarsening of the velocity field.

We next examine how the defect density evolves during the coarsening process. To this end, we quantify the distribution of $+1$ defect types (asters, spirals, and vortices) by computing their probability at early ($\sim 2.5$ h) and late ($\sim 5$ h) times after serum stimulation, pooling data across samples from four independent experiments to improve statistical robustness (Fig.~\ref{fig:fig2}c). Defects are classified according to their flow topology and orientation (inward- vs. outward-pointing) using the radial component of the velocity field around each defect core (see Methods). At early times, the defect population is strongly dominated by outward-pointing asters and spirals, consistent with the fact that newly nucleated defects inherit the expansion-driven migration originating from former compressive sites. At later times, this distribution shifts towards inward-pointing spiral defects, indicating a systematic reorganization of defect structure during coarsening (Supplementary video 5). This temporal evolution reflects a crossover from an initial regime in which mechanically triggered expansion centers dominate the dynamics, to a regime in which defect-mediated migration governs the restructuring of the tissue into larger, coherently migrating domains.

\begin{figure}[t!]
    \centering
    \includegraphics[width=\linewidth]{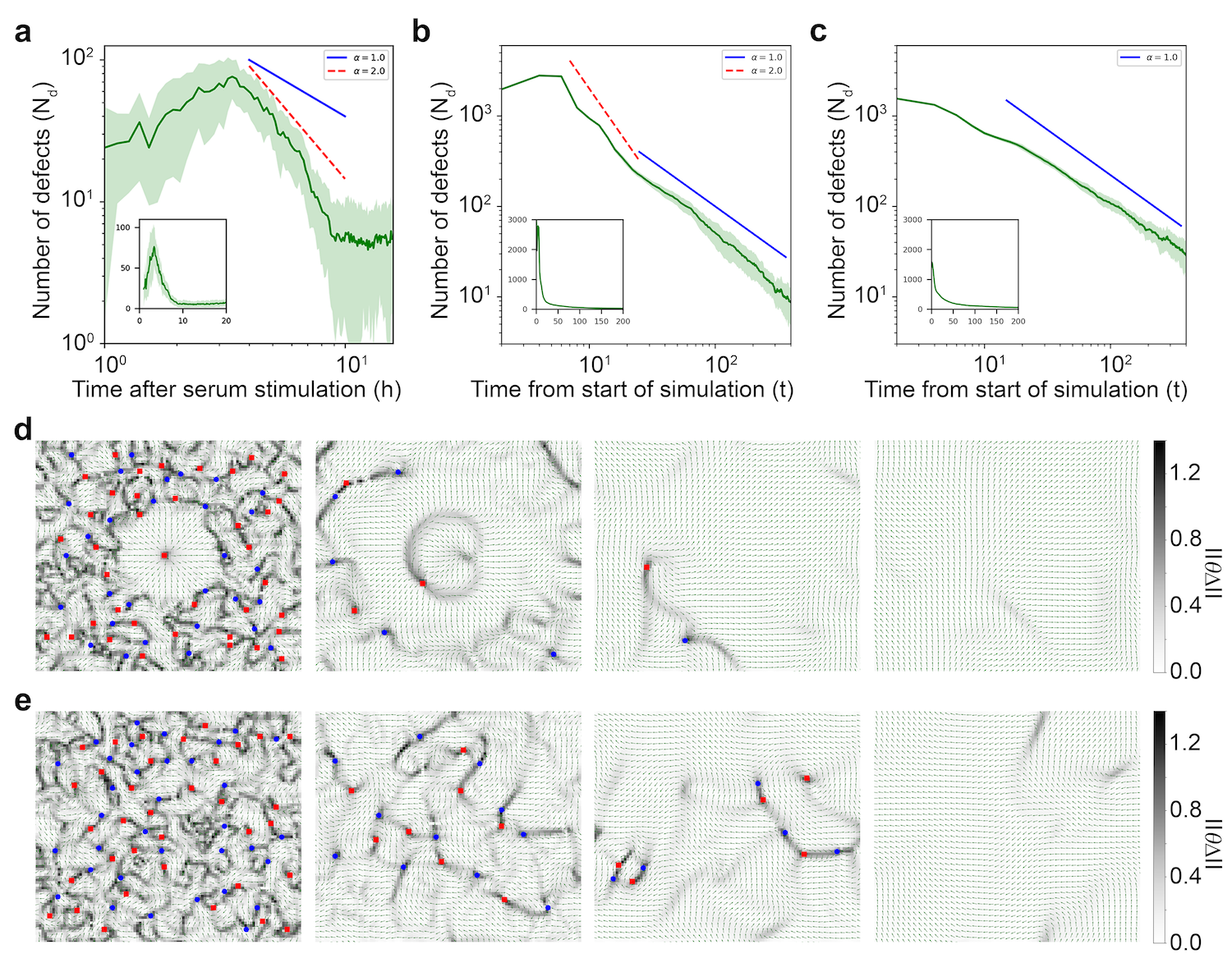}
    \caption{\textbf{Scaling laws for defect coarsening in experimental and simulated data}. {\bf{a-c}}, Scaling laws of the total number of topological defects $N_d(t)\sim t^{-\alpha}$ during defect coarsening dynamics. Log-log scale in the main figures, and linear scale in the insets. {\bf{a}}, Experimental data, from one representative experiment, with $\alpha\approx 2$. {\bf{b}}, Numerical simulations with multiple initial actomyosin expansion centers corresponding to $\alpha\approx 2$ in early times and a cross-over to $\alpha\approx 1$ in the late times (simulation parameters: $\Pi = 0.0105$, $\varepsilon = 0.08$, $D = 1$, $\tilde{v_0} = 0.025$). {\bf{c}}, Numerical simulations with suppressed actomyosin dynamics ($\Pi=0$, and simulation parameters as in (b)) corresponding to $\alpha\approx 1$.
    {\bf{d}}, Snapshots from numerical simulations with a single initial actomyosin expansion center. The magnitude of the phase gradient $\|\nabla\theta\|$ showing the domain walls. Topological defects are marked by the blue circles ($q=-1$ charge) and red square ($q=+1$ charge). {\bf{e}}, Snapshots from numerical simulations without initial actomyosin expansion center. The magnitude of the phase gradient $\|\nabla\theta\|$ showing the domain walls. Topological defects are marked by the blue circles ($q=-1$ charge) and red square ($q=+1$ charge).}
    \label{fig:fig3}
\end{figure}

\subsection{Fast defect coarsening dynamics}
The experimental observations above suggest that asters nucleating at pre-existing compressive stress sites play an important role in the defect coarsening dynamics. To quantify their impact on the phase-ordering kinetics, we track the total number of topological defects, $N_d(t)$, as function of time following serum stimulation. 

As shown in Fig.~\ref{fig:fig3}a and supplementary figure S1, the dynamics exhibit two distinct regimes. Immediately after stimulation, the system undergoes a rapid proliferation phase in which the total number of defects increases dramatically as the expanding asters nucleate domain walls and pairs of $\pm 1$ defects. This regime is followed by a defect coarsening phase characterized by an algebraic decay \(
N_d(t)\sim t^{-\alpha}\). Interestingly, we measure a scaling exponent $\alpha \approx 2$, faster than the classical $t^{-1}$ scaling expected for coarsening governed by pairwise defect interactions~\cite{yurke1993coarsening,chardac2021topology}. In classical phase ordering kinetics, oppositely charged defects interact through long-ranged Coulomb-like forces, $F(d)\sim s_1s_2/d$, leading to the diffusive growth of ordered domains and the scaling relation $N_d(t)\sim t^{-1}$. The experimentally observed exponent therefore indicates that defect dynamics in epithelial monolayers are not governed solely by pairwise elastic interactions.

Instead, our experiments reveal a coarsening regime dominated by advective transport. Mechanically preconditioned compressive sites reorganize upon serum stimulation into transient extensile asters that generate persistent collective migration, which advects both topological defects and polarity domain walls, thereby accelerating annihilation dynamics beyond the limit set by pairwise defect interactions. The observed scaling $\alpha \approx 2$ is therefore consistent with a coarsening mechanism governed by advection driven by stress-induced collective migration.

\subsection{Active elastic model with stress amplification}
\label{Sec:AES_model}

To test whether preconditioned active stress heterogeneity is sufficient to produce accelerated defect coarsening, we extend the active elastic model (AEM)~\cite{baconnier2025self,laang2024topology,angheluta2025topological} by introducing a positive feedback between tissue deformation and actomyosin stress amplification. The model describes self-propelled cells arranged on a deformable elastic lattice, coupled through polarity self-alignment and elastic interactions.

In the microscopic formulation, each cell $i$ is characterized by its position $\mathbf r_i$, polarity $\mathbf e_i$, and local actomyosin concentration $c_i$, for $i=1,\ldots,N$. The dimensionless equations of motion are
\begin{eqnarray}
\label{eq:microscopic_dimless}
\begin{cases}
 \dot{\mathbf r}_i = \sum\limits_{j\in\partial i} (|\mathbf r_{ij}| - 1)\hat{\mathbf r}_{ij} - \Pi \sum\limits_{j\in\partial i} (c_j - c_i)\hat{\mathbf r}_{ij} + \tilde v_0 \mathbf e_i, \\
\dot{\mathbf e}_i = \chi (\mathbf e_i \times \dot{\mathbf r}_i)\times \mathbf e_i,\\
\dot c_i = - \varepsilon c_i + \sum\limits_{j\in\partial i} \hat{\mathbf r}_{ij}\cdot (\mathbf r_{ji} - \mathbf r^{eq}_{ij}) + D \sum\limits_{j\in\partial i} (c_j - c_i),
\end{cases}
\end{eqnarray}
where $\hat{\mathbf r}_{ij} = (\mathbf r_i - \mathbf r_j)/|\mathbf r_i - \mathbf r_j|$. Here, $\Pi$ controls the strength of mechanochemical feedback between deformation and active stress, $\varepsilon$ is the actomyosin turnover rate, $D$ the diffusivity, $\tilde v_0$ the self-propulsion speed, and $\chi$ the polarity alignment rate (see SI for derivation). The polarity field governs collective migration and defect dynamics, while the elastic network mediates stress transmission across the tissue.

In the continuum limit, for small displacements $\mathbf u_i = \mathbf r_i - \mathbf r_i^{eq}$ with $|\mathbf u_i|\ll l_0$, the microscopic dynamics coarse-grain to an active viscoelastic gel,
\begin{subequations}\label{eq:twofield}
\begin{align}
\partial_t \mathbf u &= \mu\nabla^2\mathbf u + (\lambda+\mu)\nabla(\nabla\!\cdot\mathbf u) + \Pi\nabla c + v_0\mathbf p, \label{eq:u_lin}\\
\partial_t c &= -\varepsilon c - \nabla\!\cdot\mathbf u + D\nabla^2 c, \label{eq:c_lin}
\end{align}
\end{subequations}
consistent with previous formulations~\cite{laang2022mechanical} (detailed derivation in the SI). Within this framework, heterogeneous actomyosin concentrations induce local tissue deformations that generate compressional and shear modes. In the nonlinear regime, these deformations give rise to propagating polarity domain walls that nucleate and transport topological defects.

Our model quantitatively reproduces the main experimental observations. As shown in Fig.~\ref{fig:fig3}d and Supplementary video 6, for finite mechanochemical feedback ($\Pi \neq 0$), localized actomyosin fluctuations give rise to metastable extensile asters surrounded by propagating domain walls. These structures organize the surrounding cell migration and act as sources of defect nucleation, advection, and annihilation, in agreement with experimental observations from Fig.~\ref{fig:fig2}a–b. When simulations are initialized with multiple actomyosin-enriched nodes, the model reproduces the defect kinetics, including the initial proliferation phase and the subsequent accelerated coarsening characterized by an exponent $\alpha \approx 2$ at early times (Fig.~\ref{fig:fig3}b).

To isolate the role of mechanochemical coupling, we suppress feedback by setting $\Pi = 0$. In this limit, actomyosin heterogeneity does not amplify, metastable asters do not form, and defect motion is governed predominantly by pairwise elastic interactions mediated by the deformation field (Fig.~\ref{fig:fig3}e and Supplementary video 7). Consistently, the coarsening dynamics follows the classical scaling law $N_d(t)\sim t^{-1}$ (Fig.~\ref{fig:fig3}c) corresponding to pairwise interactions. 

At late times, even for finite $\Pi$, the initially amplified stress heterogeneity progressively relax. As metastable asters unpin and annihilate with neighboring $-1$ defects, the advective source weakens and the system crosses over from advection-dominated dynamics to an elastic-interaction regime, recovering the asymptotic $t^{-1}$ scaling (Fig.~\ref{fig:fig3}b). 

Thus, our results establish that defect coarsening in epithelial monolayers arises from a competition between elastic relaxation and active stress amplification driven by mechanochemical feedback. The model shows that mechanically inherited stress heterogeneity generates a metastable migration pattern that advects defects and transiently enhances coarsening beyond the classical phase-ordering behavior, providing a quantitative framework that captures both the kinetics and the scaling regimes observed experimentally.

\begin{figure}[t]
    \centering
    \includegraphics[width=0.6\linewidth]{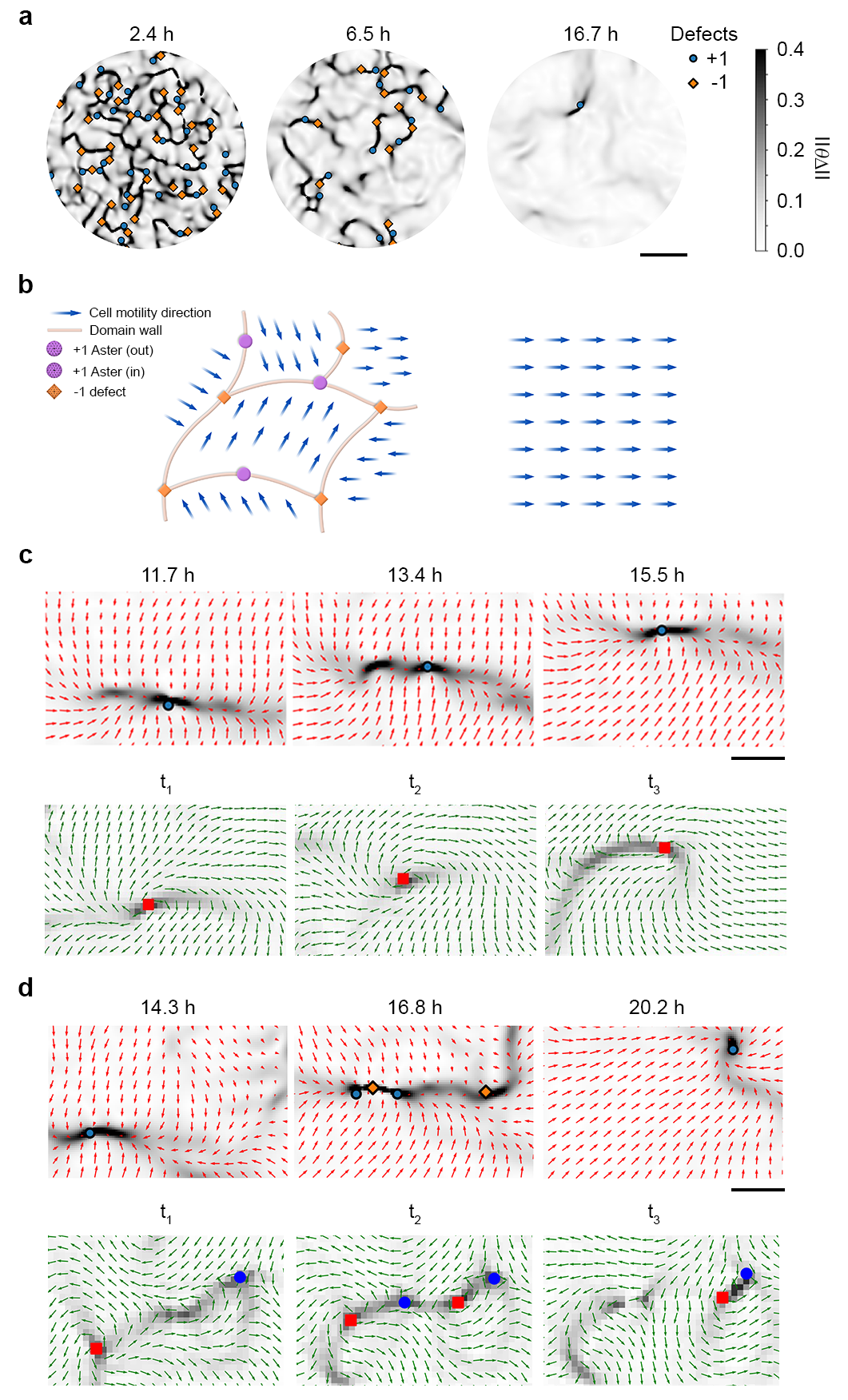}
    \caption{\textbf{Dynamics of topological defects and domain walls during defect coarsening.} \textbf{a},
Experimental snapshots of networks of domain walls laden with $\pm 1$defects during different stages of
polar ordering. Scale bar, 500$\mu$m. \textbf{b}, Graphic illustration of the polar ordering. \textbf{c}, Snapshots of the
$\pm 1$ defects interacting along domain walls (gray scale) in the normal mode for experiments (upper
panel) and model simulations (lower panel). The velocity field is represented by red arrows. Scale bar,
500$\mu$m. \textbf{d}, Snapshots of the $\pm 1$ defects interacting along domain walls (gray scale) in the sliding mode
for experiments (upper panel) and model simulations (lower panel). The velocity field is represented
by red arrows. Scale bar, 500$\mu$m.}
    \label{fig:fig4}
\end{figure}

\subsection{Domain walls as conduits for defect dynamics}
Both experiments and model simulations reveal a coexistence between polarity domain walls and topological defects during epithelial flocking. We identify domain walls as localized regions of large orientation gradients in the velocity field, quantified by $\|\nabla\theta\|$, where the local orientation angle $\theta$ is defined through the unit vector $\mathbf n=(\cos\theta,\sin\theta)$ and $\nabla\theta = n_i^\perp \nabla n_i$. In coherently migrating regions, $\|\nabla\theta\|\approx0$, whereas elevated values mark interfaces that separate domains of distinct migration direction (Fig.~\ref{fig:fig4}a). 

Figure~\ref{fig:fig4}a and Supplementary video 8 shows the dynamics of the evolving domain wall network during defect coarsening and polar ordering. At early times, the tissue is partitioned into multiple migrating domains separated by extended domain walls populated with $\pm1$ defects. The domain walls coarsen gradually and the remaining ordered domains grow in size, giving rise to large-scale coherent migration. The corresponding schematic in Fig.~\ref{fig:fig4}b illustrates how domain walls dynamically reorganize the topology of the velocity field during this ordering process.

Topological defects remain strongly localized along domain walls, which act as conduits for defect transport and interaction (Fig.~\ref{fig:fig4}a,c,d and Supplementary video 8). We identify two dominant modes of defect motion. In the normal mode, defects move together with propagating domain walls (Supplementary video 9). In the sliding mode, defects move along the domain wall, during rapid interactions including pair nucleation and annihilation events (Supplementary video 10). Figure~\ref{fig:fig4}c and d show representative experimental snapshots of such defect interactions along domain walls (upper panels) and demonstrate that the active elastic model quantitatively reproduces the same dynamics (lower panels).

This sliding motion enables defects to relocate over large distances on timescales much shorter than expected from pairwise elastic interactions alone. Because the surrounding migration pattern is strongly constrained by the position and charge of defects, their transport along domain walls continuously reshapes the global velocity orientation promoting rapid reorientation of collective migration patterns. Domain walls therefore do not merely separate ordered regions passively. Rather, they actively mediate topological transport and defect reactions that drive large-scale self-organization of the epithelial monolayer. 

\section{Discussion and conclusions}\label{Sec:Concl}
We have shown that the transition from quiescence to collective migration in epithelial monolayers is governed by pre-existing heterogeneous stress developed during the quiescent phase due to the absence of serum-born components. Upon serum stimulation, these contractile sites reorganize into metastable extensile asters that generate coordinated collective migration. These migration patterns nucleate propagating polarity domain walls and advect surrounding $\pm1$ defects, thereby driving rapid topological reorganization of the velocity field. As locally stored active stresses relax, metastable asters unpin and subsequently annihilate with neighboring $-1$ defects, resulting in the formation of patched coherent migration. 

Our results further reveal that polarity domain walls are not passive interfaces, but active dynamical structures that mediate defect motion and reaction. Defects remain strongly localized along these walls, where they undergo nucleation, advection, and annihilation events. Through this mechanism, domain walls act as conduits that reorganize the topology of the migration field over large distances and short timescales. The resulting defect dynamics give rise to a transient coarsening regime characterized by accelerated scaling, $N_d(t)\sim t^{-2}$, in contrast to the classical $t^{-1}$ behavior expected when defect motion is governed predominantly by pairwise interactions. Our experiments and model simulations therefore identify advective defect transport generated by mechanically preconditioned active stresses established during quiescence as the physical origin of the anomalously fast ordering dynamics.

The augmented active elastic model introduced here quantitatively recapitulates the experimental observations, capturing the emergence of actomyosin-induced extensile asters, the formation and propagation of domain walls that act as conduits for topological defects, and the crossover from early-time accelerated coarsening to late-time elastic relaxation. By isolating the role of mechanochemical feedback, we demonstrate that mechanically inherited stress heterogeneities are sufficient to generate the collective migration responsible for accelerated defect annihilation and rapid polar ordering.

Unlike current flocking theories, where ordering emerges primarily through alignment interactions, or jamming-unjamming scenarios, where collective motion is associated with fluidization and enhanced rearrangements, here the onset of migration is pre-patterned by internally stored mechanical stresses. The stress landscape determines where extensile asters form, where domain walls propagate, and how topological defects reorganize the tissue into ordered migrating domains.

These results demonstrate that epithelial tissues can use mechanical memory to coordinate collective migration while remaining mechanically cohesive, which has important implications for morphogenesis, wound healing, and cancer invasion.

\section{Methods}\label{Sec:Methods}

\subsection{Activation of quiescence-dependent self-ordering}
The experimental protocol used to induce quiescence-dependent activation of collective migration has been described previously~\cite{laang2022mechanical,laang2018coordinated}. Immortalized human keratinocytes (HaCaT) expressing mCherry-labeled Histone H2B~\cite{10.1083/jcb.106.3.761,laang2018coordinated} were cultured under standard culture conditions in Iscove’s modified Dulbecco’s medium (IMDM; MedProbe) supplemented with 10\% fetal bovine serum (FBS; Thermo Fisher Scientific) and 90 U/ml penicillin/streptomycin (PenStrep; Lonza). Cells were seeded in either 96-well plates (Greiner Sensoplate; M4187-16EA; Merck) at a concentration of 75.000 cells per well, or in 12-well plates (P12G-1.5-14-F; MatTek Corporation) at a concentration of 600.000 cells per well. Each well was coated with 20 µg/ml collagen IV (C7521; Merck) prior to seeding the cells. Cells were then cultured in normal growth medium at 37°C and 5\% CO\textsubscript{2} overnight. To induce quiescence, cells were serum deprived for 48 - 72 hours (h) before being stimulated with normal growth medium containing 15\% FBS.

\subsection{Live cell imaging}

\paragraph{Experimental setup I for long-term serum deprivation and serum stimulation in 96-well plates.} Image acquisition was performed using the ImageXpress Micro Confocal High-Content microscope controlled by MetaXpress 6 software (Molecular Devices). Widefield images were acquired using a 4X/0.2 NA PLAPO air objective at pixel binning 2, a filter set for the detection of mCherry fluorescence, and an environmental control gasket that maintains 37°C and 5\% CO\textsubscript{2}. Four tiled images were acquired per well every 4 or 8 minutes (min) for 30 to 50 h. The acquired time series were then processed by stitching to generate movies that display cell migration patterns on the entire well bottom surface. In this setup, image acquisition began 1 h after serum stimulation.

\paragraph{Experimental setup II for short-term serum deprivation and serum stimulation in 12-well plates.} Migration patterns were recorded using a Zeiss Axio Observer.Z1 widefield microscope controlled by ZEN blue software (Zeiss). Widefield images were acquired with a 10X/0.3 NA Plan NEOFLUAR air objective at 37°C and 5\% CO\textsubscript{2}. Images were collected every 16 min, and at each time point both the mCherry and phase-contrast channels were acquired. For each experiment, multiple single frames were recorded from multiple wells. In these experiments, cells were imaged during short-term serum deprivation for 8 h, and imaging was continued at the same positions immediately following serum stimulation.

\subsection{Velocity field reconstruction and topological defect analysis}

Velocity fields $\mathbf{V}(\mathbf{r},t)$ is reconstructed from nuclear displacements using PIV applied to time-lapse images as previously described~\cite{laang2022mechanical}. The local velocity orientation is defined as $\theta(\mathbf{r},t) = \arg(\mathbf{V})$. Topological defects are tracked using the Halperin–Mazenko formalism~\cite{skogvoll2023unified}, which locates zeros of the coarse-grained velocity field associated with undefined orientation~\cite{angheluta2025topological}. The defect density is given by
\[
D(\mathbf{r},t) = \det(\nabla \mathbf{V})
= \partial_x V_x \, \partial_y V_y
- \partial_x V_y \, \partial_y V_x.
\]
This quantity vanishes in uniformly ordered regions and is localized at $\pm 1$ defects, with its sign determining the topological charge. Defect core positions are extracted from local extrema of $D$. The total defect number $N_d(t)$ is computed by counting spatially localized extrema of $D$. 

Defect motion can be tracked instantaneously using the topological conservation law
\[
\partial_t D + \nabla \cdot \mathbf{J}_d = 0,
\]
with conservative defect current
\[
\mathbf{J}_d =
\partial_t V_y \, \nabla^\perp V_x
-
\partial_t V_x \, \nabla^\perp V_y,
\]
where $\nabla^\perp = (\partial_y, -\partial_x)$. The instantaneous defect velocity is defined as $\mathbf{v}_d = \mathbf{J}_d / D$, evaluated at defect cores. All statistics are computed as ensemble averages over independent monolayers.

\subsection{Numerical simulations}

The microscopic model is simulated on a triangular lattice with periodic boundary conditions using an explicit Euler integration scheme. Each cell is assigned an initial polarity drawn from a uniform random distribution. For the main analysis in Fig.~\ref{fig:fig3}, simulations are performed with dimensionless parameters $\Pi = 0.0105$, $\varepsilon = 0.08$, $D = 1$, and $\tilde v_0 = 0.025$. The system size is $N = 102{,}400$ cells, and results are averaged over 20 independent realization. To model mechanically preconditioned states, simulations are initialized with spatially localized actomyosin fields of the form
$c_i = 0.3 \exp(-d^2/3)$, where $d$ is the distance from the center of each seed. These initial seeds are distributed on a regular grid with spacing of $10$ lattice units, with small random perturbations introduced to avoid identical initial conditions across realizations. At late times, finite-size effects lead to deviations from power-law scaling due to the decreasing number of remaining defects and eventual extinction in finite domains.

\subsection{Domain walls, defect morphology, and cell density}

Domain walls are defined as regions of strong spatial variation in the velocity orientation field $\theta(\mathbf{r},t) = \arg(\mathbf{V})$. The orientation field is represented as $\mathbf{n} = (\cos\theta, \sin\theta)$, and local reorientation is quantified by
\[
\nabla \theta = \mathbf{n}^\perp \cdot \nabla \mathbf{n},
\]
where $\mathbf{n}^\perp = (-n_y, n_x)$. The magnitude $\|\nabla \theta\|$ is used as a measure of local angular distortion. Domain walls are identified as regions where $\|\nabla \theta\|$ exceeds a threshold defined as two standard deviations above the mean. Defect trajectories are decomposed relative to the nearest domain wall into (i) normal motion, perpendicular to the local wall orientation, and (ii) transverse motion, parallel to the wall. For each $+1$ defect, the surrounding velocity field is averaged within a circular region of radius $R_c \sim 50~\mu$m to quantify radial and tangential flow components.

Cell density fields are computed from nuclear positions using kernel density estimation,
\[
\rho(\mathbf{r},t) =
\sum_i \frac{1}{2\pi\sigma^2}
\exp\!\left[-\frac{|\mathbf{r}-\mathbf{r}_i(t)|^2}{2\sigma^2}\right],
\]
where $\sigma$ is chosen to match the mean nearest-neighbor spacing. Density fields were analyzed jointly with defect positions and domain wall networks to quantify correlations between topology, flow structure, and mass redistribution. All measurements are averaged over $n$ independent experiments. Temporal evolution of defect statistics, domain wall networks, and density fluctuations are analyzed as a function of time after serum stimulation.

\bibliography{refs}

\end{document}